# What does it mean to think like a physicist? Insights from physics graduate students

Apekshya Ghimire * and Chandralekha Singh

Department of Physics and Astronomy, University of Pittsburgh, Pittsburgh, PA, USA

*Correspondence: apg61@pitt.edu

**Abstract**

Learning to think like a physicist (LTP) is often cited as a central goal of graduate physics education, yet what this means in practice and the extent to which physics graduate education prepares students to develop LTP and view LTP as valuable to their research and teaching remain unclear. This interview-based study, conducted with seven physics graduate students at one U.S. public research university, explores how students define thinking like a physicist and how their coursework and research experiences correlate with this development. Students emphasized that physics uniquely requires integrating physical and mathematical concepts in ways that go beyond other science disciplines. Our findings show that physics core courses, particularly electricity and magnetism, frequently emphasize mathematical techniques and content coverage at a rapid pace at the expense of deeper conceptual engagement and development of LTP. In contrast, physics elective courses and research experiences were more synergistic with and effective in fostering conceptual understanding, problem-solving skills, and identity development as physicists. Because graduate students simultaneously take core courses, conduct research, and teach introductory physics, their perspectives on LTP are particularly valuable in how physics departments may consider transforming their preparation. Their voices highlight how this transformative stage of training can either support or hinder the development of physicist thinking. Because these perspectives reflect students' experiences within a single institutional context, the findings should be interpreted accordingly. Overall, the findings, interpreted through a framework that considers decision-making, conceptual and mathematical integration, epistemological sophistication and disciplinary expectations including social and cultural norms, suggest that supporting LTP requires greater alignment between graduate coursework, research and authentic problem-solving.

# 1. Introduction and Framework

## 1.1 Framework for development of expertise in physics

In physics education, the phrase *"thinking like a physicist"* is often invoked as an essential goal especially in the context of advanced training of physics majors and graduate students, yet its meaning remains underexplored [1-5]. For some, it reflects the ability to apply mathematical formalism to solve complex physics problems, while for others it emphasizes conceptual reasoning, epistemological sophistication, and a physicist like stance toward integration of physics and math for problem-solving [6-18]. The tension in integrating conceptual and quantitative reasoning in physics problem-solving has been well documented: students often perceive conceptual reasoning as intuitive or using their "gut feeling" or "guessing" [4, 10, 19-21], whereas quantitative reasoning is viewed as involving plug-and-chug methods [22], particularly due to the way physics courses are structured. Yet, authentic physics practice demands an integration of physics and mathematics concepts in a coherent seamless manner and becoming epistemologically sophisticated to solve problems.

The interplay between mathematics and physics has been highlighted extensively in both physics education research and the philosophy of science. Tzanakis [6] argues that mathematics should not be viewed merely as a tool for calculation in physics, but as an integral part of the discipline's very structure. He describes the relationship as reciprocal: physics depends on mathematics to articulate its ideas, while mathematics gains meaning through its application to physical contexts. Along similar lines, Redish et al. [23, 24] emphasize that the mathematics used in physics differs from what students encounter in traditional math courses, since equations and symbols derive their significance from the physical concepts they embody, rather than from abstract mathematical rules alone. Pietrocola [17] further characterizes mathematics as the "structural language" of physics, a medium for organizing and expressing physical reasoning that transcends mere computation. Expanding on these perspectives, Uhden and colleagues [7, 8] developed a framework that identifies two distinct but connected roles of mathematics in physics: the technical role, in which math is used to perform calculations and solve equations, and the structural role, in which math supports modeling of physical phenomena, organizes relationships between quantities, and provides a scaffold for reasoning.

This integrative nature of physics with mathematics distinguishes it from many other scientific disciplines [18, 25-31]. Thinking like a physicist requires not only facility with advanced mathematics but also the ability to connect equations to physical meaning, interpret results within conceptual frameworks, and flexibly shift between formal and intuitive reasoning [8, 25, 26, 32-38]. Although prior literature often contrasts conceptual and quantitative reasoning, several scholars have argued that physicist thinking is better understood as a dynamic interplay between the two. In our study, we use the terms conceptual and quantitative problem-solving noting that this is the prevalent language in physics education and is widely used by instructors, students and researchers. We recognize that these categories are fundamentally inseparable, aligning with Modir et al.'s framework [39], which distinguishes between different dimensions of reasoning in physics. Conceptual physics and conceptual mathematics reflect understanding of underlying principles, algorithmic mathematics represents the technical component of mathematical problem-solving, and algorithmic physics often involves memorization of facts and standard procedures [39]. Instructors often agree on the importance of developing these skills, but students' experiences often suggest that physics courses can overemphasize and reward technical procedures at the expense of conceptual foundation and building of an expert-like knowledge schema [31, 40-50]. As a result, students may struggle to recognize what is important in the course material they are learning, limiting their ability to integrate knowledge from their physics problem-solving into their schema and building robust expert-like schema.

Graduate students in physics occupy a particularly important position within this discussion. They are simultaneously learners in demanding physics core courses, active researchers confronting unsolved problems, and in many cases, teachers tasked with helping undergraduates develop their own physicist-like thinking. Their perspectives can illuminate how formal instruction, research and teaching intersect to shape and are shaped by the development of professional identity as physicists. Prior studies in physics education have largely focused on undergraduates' problem solving and epistemological beliefs, and fewer studies have investigated graduate students' views of what it means to think like a physicist and how their educational trajectories foster or hinder this development [5, 51-56].

This study addresses that gap by examining how U.S. graduate students define, experience and value "thinking like a physicist". Using semi-structured interviews, we explored how students described their coursework, research, and teaching as contributing to their development as physicists and vice versa. Particular attention is given to how they perceive the role of conceptual understanding, mathematical reasoning, pacing of their course materials, and instructional practices in shaping their thinking like a physicist. By centering the voices of graduate students, this work contributes to a deeper understanding of how physicist-like thinking is cultivated and harnessed, the challenges students encounter, and the implications for designing graduate-level instruction that better integrates physics and math, opportunities for epistemological sophistication supporting the holistic development of future physicists.

Because this study is situated within the United States, it is important to clarify that the structure of graduate education in the U.S. differs substantially from many other countries, particularly European systems in which graduate students often refer to master's students rather than PhD students. In the U.S. context, graduate physics students typically enter PhD programs after completing a bachelor's degree and spend their early years taking graduate-level core courses and earning a master's degree along the way before transitioning more fully into research. We situate the study within this structure to help readers better understand what stage of training the participants were in and how they were engaging with disciplinary practices at the time of the interviews.

Our analysis revealed that graduate students' perspectives on learning to think like a physicist (LTP) were multifaceted, encompassing not only the integration of conceptual and quantitative problem-solving but also issues of identity, epistemology, and pedagogy. Most participants described tensions between traditional instructional practices in graduate core courses, often fast-paced and mathematically focused and their own needs for conceptual clarity and need to build schema like expert physicists, time to reflect, and opportunities for discussion. However, elective courses, research experiences, and teaching responsibilities were often seen as more effective spaces for cultivating physicist-like thinking and harnessing it.

### 1.2 Analytical Framework

Developing the ability to "think like a physicist" is a multi-faceted process [4, 5] that involves the integration of conceptual reasoning, mathematical modeling, epistemic practices and socialization into the norms and values of the physics community. Prior work on thinking like a physicist explores the epistemic tools and reasoning strategies that physicists use when approaching complex problems [15, 57, 58]. For example, Price et al. [15] characterize authentic scientific thinking through practices such as making justified approximations, analyzing limiting cases and shifting between mathematical and conceptual representations. These practices help experts to determine what is relevant, what can be ignored and how to frame problems in productive ways. These researchers [15] identify 29 distinct decisions that can be implemented in the problem-solving process, where the goal is not merely arriving at an answer but engaging fully with the reasoning process itself. Building on this work, Robbins and Burkholder examine expert decision-making in graduate-level physics contexts [57, 58]. In their study in graduate courses, they identify the specific decisions that experts and advanced students face such as selecting representations, choosing solution methods, making assumptions and evaluating the plausibility of intermediate results. Together, these studies highlight that physicist thinking is defined not only by conceptual and mathematical fluency but also by the ability to coordinate decisions about models, representations and solution strategies.

Thinking like a physicist also involves navigating the social and cultural expectations of the discipline. Ulriksen argues that physics departments and their culture often implicitly construct an "implied student" whose motivations, background and ways of thinking align with expected disciplinary norms in the established physics culture [59]. In Ulriksen's definition, the implied student could be understood as "the study practice, the attitudes, interpretations and behavior of the student, that is presupposed by the way the study [38] is organized, the mode of teaching and assessment, by the teachers and in the relations between the students, enabling the students to actualize the study in a meaningful way. It is presupposed that students can act in and with this structure, and it provides the students with specific possibilities for acting in the study." Students whose experiences diverge from these implicit expectations may feel less recognized or supported. Our analytic framework aligns with Ulriksen's emphasis on a holistic, enculturation-based perspective, in which becoming a physicist is not simply a matter of completing coursework or earning grades, but of being gradually inducted into the social, cultural and intellectual environment of physics [38]. From this perspective, students' experiences, interactions and sense of belonging all shape how they come to understand what counts as being a physicist. This framework foregrounds the tension between the norms and expectations imposed, e.g., by graduate or undergraduate physics programs and how students navigate, resist or internalize these expectations to be viewed as successful physicists. This perspective is particularly useful for understanding how our graduate student participants interpreted disciplinary expectations and how they navigated, negotiated or internalized these expectations.

Together, these frameworks conceptualize thinking like a physicist as a cognitive, epistemic and social process. They suggest that physicist thinking is cultivated not only through conceptual and mathematical reasoning and expert-like decision-making, and engagement in authentic disciplinary practices [57, 58], but also with gradual socialization and pressure to yield to the norms and develop identities viewed as desirable in the physics community [38]. These frameworks and perspectives guide our interpretation of graduate students' reflections. We examine their descriptions of what helps and hinders their development of LTP not simply as individual viewpoints but as insights into how graduate training structures result in opportunities and constraints for cultivating the physics habits of mind, reasoning strategies and identities valued in physics.

## 2. Research Questions

This study was guided by three overarching research questions, each designed to capture different aspects of how graduate students develop and harness the ability to "think like a physicist". The following research questions frame our analysis of graduate students' perspectives on LTP in interviews, situating their experiences in developing and harnessing LTP across three broad themes, personal development, graduate coursework and professional preparation beyond the coursework:

RQ1. How do graduate students define and develop the ability to think like a physicist, and how do they perceive their views evolving over time?

RQ2. What role do graduate courses play in shaping graduate students' ability to think like a physicist?

RQ3. How do graduate students perceive the importance of thinking like a physicist for research and teaching?

## 3. Methodology

### 3.1 Geographic and Educational Context

This study was conducted in the United States where the structure and expectations of graduate education differ significantly from those in many other countries. The present study was carried out at a large public research university with a total enrollment of roughly 35,000 students, including about 25,000 undergraduates and around 10,000 graduate students across its campuses, providing a large and research-intensive environment. In the U.S. system, graduate students in physics are typically admitted into integrated master's and Ph.D. programs, with most students accepted only if they intend to pursue a Ph.D. While enrolled, they receive full tuition coverage and a monthly stipend in exchange for research or teaching responsibilities.

Graduate programs generally require students to complete a combination of core and elective courses during the first two years. In the department where this study was conducted, a typical graduate physics class size ranges from about 20 to 30 students. The students typically fulfill program requirements by taking either four core courses and five electives or five core courses and four electives, depending on their interests and advising from Ph.D. supervisor. In this department, the mandatory core courses include Quantum Mechanics, Electricity and Magnetism, and Statistical Mechanics, with students selecting their remaining core requirement from either Dynamical Systems or Quantum Mechanics II. During this period, graduate students often serve as teaching assistants and are expected to pass qualifying examinations before transitioning into full-time research. Because the participants in this study were drawn from this U.S. Ph.D. context, their perspectives on "thinking like a physicist" reflect the experiences of students who have already completed substantial upper-level and graduate-level physics coursework and who are positioned at an intermediate stage of their doctoral training.

### 3.2 Participants

We conducted semi-structured interviews with seven graduate students from one physics department. The participants were volunteers who varied in the research area and year in their program: two were doing Ph.D. in astrophysics (one theory, one observational), one in particle theory and four with emphasis in physics education research. Of the seven participants, four identified as women and three as men. Two participants earned their undergraduate degrees outside the United States from Asian countries. The other five participants were US citizens (four were White and one was Hispanic). All participants had successfully completed the comprehensive examinations required for the PhD program and had served as graduate teaching assistants for at least one semester. At the time of the interviews, three were in their second year, one was in their third year, two were in their fourth year, and one was in their sixth year of the PhD program that typically spans six years on average (including master's along the way). Table 1 provides detailed information about each participant.

**Table 1**. Relevant information about the interview participants.

| Participant | Gender | Year | Research Area | US Undergraduate |
|---|---|---|---|---|
| Brianna | Woman | 2nd | Physics Education Research | Yes |
| Noah | Man | 2nd | Physics Education Research | Yes |
| Grace | Woman | 2nd | Physics Education Research | No |
| Kylie | Woman | 3rd | Physics Education Research | No |
| Ethan | Man | 4th | Astro Observation | Yes |
| Sophie | Woman | 4th | Particle Theory | Yes |
| John | Man | 6th | Astro Theory | Yes |

Although their research areas differed spanning both physics education research (PER) and non-PER subfields, all participants were PhD candidates working toward the same doctoral degree in physics. They had each completed the same set of graduate core courses and held teaching assistant positions within the department, giving them broadly comparable academic and instructional experiences. All of the requirements for completing the physics PhD were the same for all seven interviewees. In this sense, their trajectories within the program were similar despite disciplinary differences in research focus.

Participation was entirely voluntary. Students were invited to take part in the research interviews, and those who agreed were informed that their responses, including quotations, could be included in research publications and presentations. Verbal informed consent was obtained prior to all interviews. To protect confidentiality, pseudonyms were assigned to participants, and no identifying details are reported here.

**3.3 Survey Development and Data Collection**

To investigate graduate students' perspectives on LTP, we developed and validated a set of 16 interview questions through a collaborative iterative process. Both researchers brainstormed and discussed potential areas of inquiry based upon our goals. Some of these questions were inspired by prior study on physics graduate core courses that suggested that they may be missed learning opportunities for developing a functional understanding of physics [41]. After designing these questions, they were further refined and shared with four physics professors for feedback. During this stage, faculty members reordered several questions, suggested clearer phrasing, and recommended adding a few new questions that they felt would better capture students' reasoning and experiences. In this way, the questions were strengthened and validated through a process of repeated refinement and discussion before being finalized.

The interview protocol consisted of 16 structured questions designed to probe students' experiences with graduate coursework, research training, and the development of their identity as physicists overall. The interview protocol primarily focused on participants' own graduate experiences with one question (Q3) explicitly asking about how physicists approach problems differently from other disciplines, which invited responses about the broader physics community. However, we did not make distinctions between personal experiences and perceptions of the wider physics community during the interviews conducted. Except in response to Q3 in the Appendix, when participants shared observations or thoughts about the physics community more broadly, they did so on their own initiative. The interview questions are provided in the Appendix. Interviews were conducted over Zoom, recorded with permission, and each lasted between 30 minutes to one hour depending on the participants' responses.

### 3.4 Transcription and Data Preparation

The interviews were automatically transcribed using Zoom's transcription tool, and both the transcripts and video recordings were retained for accuracy checks. The transcripts were first cleaned to remove filler words such as *"like", "kind of", "you know", "yeah"* and other repetitive phrases that did not add meaning. In cases where Zoom produced incorrect transcriptions, we returned to the video recordings to carefully review the participants' speech and correct errors. This process ensured that the integrity and intended meaning of participants' responses were preserved.

Due to glitches with the Zoom video of one participant, transcript is not available. As a result, this interview was not coded, as we did not have access to the participant's exact wording. However, the interviewer's detailed notes indicate that this participant's perspectives aligned closely with themes expressed by the other graduate students, and these notes were used only to confirm the consistency of emergent patterns rather than to generate codes.

### 3.5 Data Analysis

Following transcription and cleaning, we engaged in a systematic coding process. The analysis of interview data was guided by Saldaña's work [60] on qualitative coding. Both researchers reviewed the interview transcripts and video recordings at least once to listen carefully to what participants were saying and to familiarize themselves with the data. They then met to discuss observed patterns, participants' lines of reasoning and the experiences that recurred across interviews. These discussions were extensive and informed the decisions regarding how the data should be represented prior to formal coding.

To understand the types of coding applicable to interview data, we reviewed the coding strategies described by Saldaña and determined that structural coding was most appropriate for the first round of coding for our study. Our analytical approach is grounded in phenomenographic analysis, which aims to identify the qualitatively different ways that participants experience, understand or conceptualize a particular phenomenon [61]. This approach aligned with our interview data, which focused on how graduate students reason about and describe "thinking like a physicist". Phenomenography emphasizes capturing variations in participants' meanings and experiences rather than seeking a single unified interpretation, making it well suited to our goal of characterizing distinct ways students articulated physicist thinking [61].

Within this phenomenographic orientation, structural coding allowed us to segment the data according to the major conceptual areas explored in the interviews, creating an initial organizational scaffold for identifying categories of description. The themes and subthemes in this study emerged through structural coding, in which the interview questions were used to group participant responses into three overarching themes and their associated subthemes. They were therefore directly tied to the prompts asked during the interviews. In contrast, the codes were generated from recurring patterns in participants' actual responses and were grounded in the specific thoughts shared by the graduate students.

| Theme 1: How do graduate students define and develop the ability to think like a physicist, and how do they perceive their views evolving over time? ||
|---|---|
| **Subthemes** | **Codes** |
| Definition and evolution | Conceptual understanding as the core |
| | Broad, flexible, and persistent problem-solving approach |
| | Prioritizing thinking over outcome |
| Process of LTP and difference from other disciplines | Learning through practice, curiosity and social environment |
| | Concept-first approach with integrated math |
| Self-perception and alignment with graduate experience | Evolving confidence in LTP |
| | Graduate courses as foundation |
| | Perceived gaps and misalignment |
| Theme 2: What role do graduate courses play in shaping students' ability to think like a physicist? ||
| **Subthemes** | **Codes** |
| Role of graduate courses in LTP | Emphasizing conceptual understanding |
| | Assessments that foster deep thinking |
| | Advanced courses enhance physicist thinking |
| Course pacing, its impact and suggested changes | Rapid pacing limits conceptual understanding |
| | Difference in levels of instructor awareness and responsiveness |
| | Reorganization of course content |
| Theme 3: How do graduate students perceive the importance of thinking like a physicist in their research and teaching? ||
| **Subthemes** | **Codes** |
| Importance of LTP for research | Critical for success in research |
| | Role in teaching and mentorship |
| Teaching goals, course content priorities, and pedagogical techniques for LTP | Modification of the standard content |
| | Use of various pedagogical techniques |
| | Connecting learning to research and real world |

Figure 1: Themes, subthemes, and codes on graduate students' perspectives on LTP.

After extensive discussions, both researchers conducted the initial round of structural coding and produced a preliminary set of themes and subthemes based on the interview questions. One researcher then produced a preliminary set of codes that reflected participants' responses. Both researchers then collaboratively reviewed these coded excerpts, refining the codes in a second round and considering alternative ways of grouping and interpreting participants' ideas. This iterative process ensured that the final coding scheme reflected both the structure of the interview protocol and the emergent patterns in participants' perspectives, capturing the depth and nuance of their experiences.

Overall, both researchers agreed upon the themes, subthemes and codes generated during the coding process. Only one disagreement arose, regarding the number of codes to use rather than the interpretation of participants' responses. This was resolved through discussion, during which code boundaries were clarified and the final set of codes was agreed upon. This process ensured that the coding process including themes, sub themes and codes (summarized in Figure 1) was representative of the data while allowing for analytic depth.

The 16 interview questions were condensed into three analytical themes which directly align with our three research questions. These themes were then divided into seven subthemes, that captured major areas of interest across the dataset [60]. Within each subtheme, participants' responses revealed patterns that were further refined into 2 to 3 codes. These codes highlighted nuances in how students conceptualized thinking like a physicist, how they perceived their coursework experiences to help them think like a physicist, how they connected these experiences to their research, teaching or professional goals and what they would do if they were teaching to help their students in LTP in the future if they became faculty.

## 4. Results and Discussion

Graduate students in physics described LTP as a complex and evolving process shaped by their coursework, research experiences, broader physics culture and teaching aspirations. Their perspectives revealed 7 interconnected subthemes that illustrate both the opportunities and challenges in cultivating and harnessing physicist-like thinking during physics graduate education.

**RQ1. How do graduate students define and develop the ability to think like a physicist, and how do they perceive their views evolving over time?**

The first theme probed students' definitions of thinking like a physicist, the processes through which they learned to do so, the evolution of their confidence, and the influence of coursework, research, peers and faculty on their development. These questions were intentionally broad and reflective, capturing the holistic and longitudinal nature of LTP as students move through graduate school. Thus, this theme represents students' personal and intellectual trajectories on how their thinking is shaped by cumulative experiences, from classroom expectations and collaborative norms to research identity formation.

4.1 Definition and evolution

This subtheme captures how graduate students define what it means to "think like a physicist" and how their understanding of this concept developed over time. Students describe shifts in perspective from undergraduate to graduate studies, highlighting the evolving nature of what counts as physicist thinking.

*4.1.1 Conceptual understanding as the core*

Four participants emphasized the centrality of conceptual understanding in thinking like a physicist.

Sophie highlighted, "When you're trying to think like a physicist, you have to really think about the concepts of the scenario before you can really apply any math to it. Math is more of a tool to get you to the final answer, as opposed to problem-solving skill. I think the concepts are more of what you need to think about first, when you're trying to solve a problem, and then the math will follow from that."

Similarly, Noah reflected on the role of concepts in structuring reasoning: "I think that the concepts are really important when you're thinking like a physicist, because I think that using those concepts to build up the base of whatever you're trying to think about or understand is really important. Physicists tend to build their reasoning or justification from physical concepts that we see or know about. And then thinking about the math, I would say, is perhaps a little less important but it can be very helpful to guide thinking about how or why something would change if something else was added in."

Brianna also reinforced the importance of conceptual grounding noting that, "Thinking like a physicist means probably…basically identifying the underlying concept that you need and then using the details from the situation or problem, or whatever it is to build on that. Going through all these graduate physics classes, I have seen just more and more sophistication of math that people use to solve physics problems. Yes, you need the math to solve the problem, but the physics part of it to me is identifying the concepts and using them appropriately."

Taken together, these accounts suggest that for some graduate students, conceptual grounding is a defining feature of physicist thinking and a mark of disciplinary expertise. This aligns with work showing that experts flexibly coordinate conceptual models with mathematical aspects to navigate problem-solving.

*4.1.2 Broad, flexible, and persistent problem-solving approach*

Three of the participants described thinking like a physicist as involving a broad, adaptable approach to problem-solving, rather than reliance on memorized methods.

Sophie explained, "I feel like thinking like a physicist has to do with being able to solve kind of any problem that's put in front of you, or at least have an idea of solving pretty much any kind of problem...thinking like a physicist has to do with having a framework or techniques that you apply to problem-solving as opposed to memorizing how to go about solving particular problems. So, I think having a broad approach to problems and not getting sucked into the details is another important aspect of thinking like a physicist."

Kylie emphasized the importance of persistence in this process: "For me, thinking like a physicist, first of all, is having persistence when you are thinking. So, a person who has the quality of being a physicist will not be giving up easily and will be more engaged in the process of thinking. It involves a person who thinks deeply about the concepts and thinks how certain phenomena can be related to the physical experience outside. They try to relate it with that and there is a good mathematical blending so it's not like you know some fact and start believing it. But you also think how mathematically it could be proven and how you can look at it being integrated physically as well as mathematically."

Thus, rather than relying on memorized knowledge and procedures (what Modir et al. [39] term algorithmic physics), these students highlighted the ability to approach unfamiliar problems with general strategies and an orientation toward sensemaking as thinking like a physicist.

*4.1.3 Prioritizing thinking over outcome*

Three participants reflected on how their understanding of what it means to think like a physicist has evolved over time, moving away from the idea of simply having the "right answers" to valuing problem-solving approaches and conceptual reasoning.

Sophie shared, "I think definitely my perception has changed. I think it has more to do with how you think about problems and how you approach them, and not necessarily just knowing the right answers to them. Because at least now that I'm almost a 5th year grad student, I've had a lot of discussions with people and met a lot of scientists who I feel don't always know the answers to things but have a hard time admitting to not knowing the answers to things. And being a physicist is really more about knowing how you could find the answer, not about having the answer...I feel better about being a physicist because I think that knowing how to solve problems and how to think about them is more interesting and more important than just having the right answers."

Grace described a similar evolution, "I have had different kinds of thinking about what this means over the years, I would say maybe initially, I thought, being a physicist is basically having all the right answers somehow from the beginning and knowing where you want to go with the problem, how to solve it and everything right at hand. And I think some parts of that still linger. It's not something that goes away because of the culture of what you typically see in the physics department. But at least for some of it, I have started to think that thinking like a physicist is not necessarily that you know the answers, but it's more like being able to make connections, even if you're making mistakes and being open to work with other people."

These reflections suggest that graduate students' self-perception as physicists shifts over time, emphasizing problem-solving strategies, conceptual understanding, collaboration, and openness to uncertainty rather than simply possessing correct answers. These reflections also describe thinking like a physicist as a process-oriented activity that involves breaking down problems, identifying principles and evaluating the intermediate steps. Rather than focusing solely on arriving at the correct answer, students emphasized an approach or perspective, echoing prior descriptions of authentic physicist practice.

## 4.2 Process of LTP and difference from other disciplines

Here we discuss students' reflections on the cognitive processes that underlie LTP, such as problem framing, abstraction, and conceptual reasoning. They also contrast these ways of thinking like a physicist with practices in other disciplines, emphasizing what makes physicist thinking distinctive.

### *4.2.1 Learning through practice, curiosity and social environment*

All participants in our study emphasized the importance of repeated practice, persistent engagement, and innate curiosity in developing the ability to think like a physicist.

Sophie reflected on how these elements interact, "I think at least from my experience, it does come from just a lot of practice for one. You can't expect to gain such broad skills without practicing a lot. But I think trying to answer questions that aren't just put in front of you, but actually asking questions about the world, and kind of just having an innate curiosity, I think is part of thinking like a physicist. I don't think you need to know how to solve problems, to at least want to ask the questions of how to solve them. So definitely just curiosity is part of it. I did a lot of problems throughout my career that I was like, okay, these are very similar to what we've seen in class. And then over the years, you eventually start seeing problems you've never seen before and going through that process of not just matching what you learn in class, but coming up with new ideas or solving problems that are new can help you think like a physicist."

Noah also highlighted the role of embracing uncertainty and learning from not knowing, "I feel like it takes a lot of practice, maybe a lot of asking…I've found being willing to not know why kind of helped me lead into that thought process of understanding…And being able to get past that weirdness of admitting that you don't know allows you to LTP and then lots of application would probably be very helpful…there's a certain problem-solving process that I developed because of the way that I think."

Grace emphasized the importance of the learning environment and role models, noting, "In my opinion, part of it is just being in the environment, being in contact with and seeing other people and having role models that you see how they work and how they approach solving problems. So that can either be your professor in a course or if you're a student, then other students who are solving problems in an efficient way. So, I think it's a process. It's not something that has a secret code to it that you can just say, this is how I can LTP....People have their own perception of what thinking like a physicist means, so it can be a different process for everyone."

Together, these accounts suggest that developing physicist like thinking involves sustained practice, an openness to new and unfamiliar problems, and a persistent curiosity that drives deeper understanding. These reflections also show that students gradually take on the expectations of how a physicist is supposed to think and work, often by practicing and learning from the people around them. In this way, they are both learning and living out the role of the "implied student" [59] as they develop their physicist thinking.

*4.2.2 Concept-first approach with integrated math*

Four of the participants emphasized that thinking like a physicist requires integrating conceptual reasoning with mathematical aspects of problem-solving, distinguishing it from other disciplines.

Sophie shared her own experience, "I just think that physicists don't memorize things quite as much as other disciplines because you can't get by with memorization if you're solving problems that you've never seen before. I generally try to take a big picture approach to a problem so that I can really conceptually understand the situation and then start to think about how to use the things I've seen before, to solve the problem in front of me as opposed to just going off of exactly what I did before. If you're used to memorizing how to do something, you might go straight to the math and rely on the math to tell you how to do the problem as opposed to relying on the concepts to tell you how to do the problem. So, I think that's a big difference with physicists. It is that if you're doing well at solving physics problems, you're usually thinking about the concepts before the math. And some of these other disciplines, may be are more intimidated by the math, [they] try to do the math part first, and I just feel you can't really do math until you understand conceptually what you're trying to do. I definitely see that in my teaching."

Similarly, Ethan reflected on his own experience but framed it within broader claims about the physics community, "I think physicists, due to the curriculum that I've been exposed to so far, we're required to have a pretty decent understanding of various disciplines, primarily mathematics. You know, subdisciplines like statistics in addition to the subfields within physics, all of which kind of add onto one another like quantum mechanics or statistical mechanics. Basically, any of the mechanics [quantum, statistical, classical] rely on a pretty good fundamental understanding of mathematics, and once you have those underlying principles, I think it's maybe a bit easier to see the bigger picture. Whereas, in contrast, when I think of biology, I had a lot of friends in undergrad[uate education] who were biologists, they often were not excited to take things like organic chemistry, or any heavily math involved courses whereas that is our bread and butter as physicists. And so, I think that's how physicists approach problems differently. We are kind of forced to have a good fundamental understanding of various sub-disciplines whereas that may not be entirely true across all of the physical sciences."

Grace and Noah reflected on how physicists in general approach problems with Grace noting, "What I see with physicists approaching problems compared to other people, for example, is that physicists typically tend to have this double approach…It's like they're always focused on the concept somehow and at the same time they're using mathematical reasoning to put those ideas or concepts into the correct form. For example, you have all of these tools that you can work with, and you can actually prove something with mathematical reasoning. So, I think that's the dual approach that they [physicists] have that makes it stronger in some sense and also more difficult."

Noah highlighted differences in priorities across fields, "I definitely see a lot more emphasis on the answer for other disciplines where, for example, a lot of times students will just ask me: what's the answer. And from my experience with other physicists, I've found that usually the normal response is, that's not really important. The actual process is really important to us, and then thinking about how we do exams, it's usually very process dependent, and it's not as important to get the right answer but more heavily weighted towards process and applying concepts correctly. So, I think that the difference would be the focus on the actual process rather than the answer that is resulting from a given problem."

These reflections collectively suggest that students see physicist thinking as characterized by conceptual framing, integration of mathematics as a representational tool and an emphasis on process over product. This aligns with research showing that experts coordinate multiple representations [16, 62-65], reason conceptually, and flexibly apply mathematical structures to novel situations.

4.3 Self-perception and alignment with graduate experience

Students assess their own progress toward thinking like a physicist and reflect on how their graduate coursework has contributed to or diverged from this development. This code captures tensions between self-perception and curricular design.

*4.3.1 Evolving confidence in LTP*

All participants reflected on the evolution of their confidence in LTP. Many grounded their sense of progress in their growing ability to make decisions about how to approach problems.

Sophie explained, "I don't expect to know the answers to everything. I think part of being a physicist is not necessarily having the answer already to all the problems that you'll ever see. But knowing that you have the skills to get a decent answer. And yeah, at this point in my career, I definitely feel like I've gained those skills and that I can try to solve any problem, and I would at least have a good idea of how to start that process. And I feel confidence has a lot to do with whether you think like a physicist or not. I think I've had enough experience solving problems and now I do really trust my ability to conceptually understand situations and then learn how to solve them from there."

Others described a tension between their own practice and an "idealized" image of what it means to think like a physicist. Grace reflected, "I think in some respects, sure, for almost 8 years I've been on physics departments. So, when I compare myself to other people in other fields, there's definitely a qualitative difference in the way that I approach problems. It's more cautious I would say, the way that I treat problem-solving, and I have an overall bigger picture that I always try to think about. But also, when I think about thinking like a physicist, I still have that, as I said, ideal image of what it means to think like a physicist. That is someone that has it all. And it's a difficult picture to get rid of. A physicist is just a genius person that has all the answers and can solve any questions. And that is still somewhere in my mind. So, I don't see myself as a physicist when I think of it that way, because I never have all of the answers."

Brianna highlighted the process-oriented nature of physics thinking, "I think it's a process. Saying I think like a physicist is saying that I am a physicist. That's hard to say, but I think I do think like a physicist. I break down the problem. I try to find the underlying concepts that we need. Maybe it's not all in a linear order of, like, first we do this, and then we do this. But yeah, I definitely think that if I were given a problem, I would identify the things that are relevant and try to get out the underlying concept that we need to invoke. So yeah, in that sense, I think I do think like a physicist, somewhat."

Ethan also stressed that it's a gradual development over a lifetime noting, "I think learning how to think is a skill that one develops throughout [one's] lifetime. I don't think there is a threshold to be able to say yes, I think like a physicist. And so maybe that threshold can be subjective and because I can't identify a specific threshold to say yes or no, I or someone else thinks like a physicist, I think that's kind of hard to say. But as a graduate student, I think I've had some experience in trying to think like a physicist. So maybe my answer is yes."

These reflections show that students feel that thinking like a physicist is a gradual and ongoing process rather than a fixed threshold. Students' views show that they are both growing and unsure as they try to measure themselves against what the field seems to expect.

*4.3.2 Graduate courses as foundation*

All the participants reflected on the role of coursework in LTP, noting that the alignment of courses with conceptual understanding and problem-solving skills varied.

Sophie observed, "I think some courses do, and some courses don't as much. If you're thinking of undergrad, a lot of the courses are not necessarily prepping you to think like a physicist, but they're giving you the beginning of that, right? Like you have to know a lot of base knowledge before you can start to think more broadly. So, those definitely do start to help you in LTP, but I think it's more grad school courses, and more upper division courses where things are more complicated. They don't look as much like what you saw in class, or what you saw on the homework. They can be more complex, different scenarios. So, tackling some of those crazy problems were definitely things that helped me to think more broadly and think more like a physicist. There was one in my undergrad where the questions would be-how many leaves does it take to cover the surface of the earth. Those kinds of problems really stretch your brain to think about how to solve vague problems. That's something you would not be able to memorize how to do, right? If you're doing an order of magnitude type problem, you can just look up equations…and you have to think about how you would get to the final answer. And then you can do some research to figure out what the numbers are. But you have to set up the equation yourself after thinking about what is even relevant for putting in an equation. So definitely, courses that are stressing the conceptual thinking about the problem before doing the math. I just find that the higher up you go in your career in physics, the more upper division advanced courses you take, the more concepts are stressed in problem solving."

Noah emphasized the role of repetition and practice alongside conceptual focus in the graduate courses, noting, "I would say, a lot of the time it is that practice that I was talking about where there's just so much repetition that you eventually have to [LTP]. But also thinking about how usually professors are, they are also emphasizing the importance of that process, so that causes me to make that shift as well. I found that as you move up in the difficulty of physics courses, it [focus on concepts] actually does become less and they care less about the concepts, because there's generally an assumption that you already know them which is kind of interesting. Thinking about the core classes that we had to take, I would say the only one that was really a lot more focused on the concepts than doing difficult math would be probably from my experience, just dynamical systems. I felt like beyond that, in the other three [graduate core courses, quantum mechanics, electricity and magnetism and statistical mechanics], it would be like [the instructor would say], okay, here's this concept really quick and now let's do some really hard math with it. And the assumption that is usually made is that the concept is one that you've seen before and now we can apply it in new ways because you know more math. I think that hinders it [process of LTP] because you don't really get that [opportunity to LTP] as much when you're focusing on the math …and you're not asking why from a conceptual standpoint."

Ethan added that the teaching style and historical context also influenced the process of LTP, stating, "Some courses helped more than others, and I think the biggest factor is the faculty teaching that course. I think some instructors can teach a course in a way that feels like I am being hazed. That does not help me in LTP. I, in fact, feel like I'm trying to survive in that course rather than taking something away from it. I think that overall, my physics courses could have helped me to think like a physicist better. Oftentimes it seems very formal, like I remember being taught this is how we can derive this statement, and this is what the statement tells us about the nature of our world. I think maybe having some more historical context as to how these ideas came about would help me to LTP, in the sense that, having a better understanding of how we, as a field, came to arrive to the conclusions that we consider state of the field would help me in navigating the different areas of research that are currently taking place."

These reflections suggest that while graduate courses can serve as an important foundation, students often feel that the curriculum assumes the traits of the "implied student" [59] who is already conceptually fluent, mathematically confident and ready to engage with advanced material rather than actively cultivating those traits with adequate support from the learning environment.

*4.3.3 Perceived gaps and misalignment*

Participants described mixed experiences regarding how well graduate coursework aligned with their goals and supported their development of LTP. Three graduate students pointed to clear misalignments between core course expectations and the skills or conceptual understanding they needed for their research.

Sophie reflected, "I think in general core courses don't align with everyone's goals. Like the Astro[physics] Ph.D., that was created recently, doesn't have to take E&M because it doesn't really align with the goals of the Astro Ph.D. But E&M doesn't really align with my kind of research goals either, everything I do is more quantum field theory…So, there are some classes that I feel like I took and never use, and they were so difficult that I don't think I actually built a lot of conceptual knowledge from them. And so, I'm not really learning as much, and it's not going to stick with me. If the math doesn't really mean anything, it's just math. It's not gonna stay in my brain the way that the concepts and the tools of problem solving really stick with my brain. So, E&M is a specific course that I feel I didn't get a lot out of, because it wasn't that conceptual to me. It was very mathy so it was just hard to know how to solve problems in that course. And then it doesn't really apply to what I'm doing. So, it didn't feel that useful."

Similarly, Ethan shared, "I would say, probably not really [graduate core courses are not helping as much in LTP]. And I think there's probably a bigger reason for it. I think that the core courses are very focused on showing that you can do hard math and then getting a good grade, [they are] less [about helping learn] new concepts or [develop] better understanding. That was my experience, at least. I mean, I've certainly learned…from those core courses. But I would say, as a whole, they maybe have to focus more on showing that you can do problems [in a meaningful way to LTP].''

However, other participants expressed positive experiences in at least some graduate classes. Grace shared, "I feel as you move on to graduate level courses, the structure is somehow that you have smaller classes, and whenever you have less people in a class, then that interaction with the professor and student to student interaction is higher compared to large enrollment classes…you can talk to people easily, the professors are more accessible. Even though the concepts are more difficult…the structure of the courses makes it easier to interact with people and be that physicist that you want to be. So yeah, for me, I was thinking [about] that interaction element, and the fact that I wanted to be able to talk to people and collaborate with other people and have that human interaction. It was definitely something that I achieved later as I progressed."

Kylie also shared her positive experience, "Yes, because most of the times, things that we learn in the graduate level courses contain more challenging problems. Based on my experience, there were different homework problems which required a lot of time. And it also required collaborative work. So, I would say, all those skills I have learned from those graduate courses. And the content that we learn itself is, I think, at a good level, which is required to understand physics in a better way."

Together, these synergistic but somewhat contrasting perspectives show that while some students experienced a clear mismatch between core course content and their goals, others found that smaller class sizes, increased interaction and collaborative environment supported their goals. These views illustrate how instructional design, course structure and perceived relevance shape how graduate students judge the value of their coursework for learning to think like physicists.

**RQ2. What role do graduate courses play in shaping students' ability to think like a physicist?**

The second theme focused specifically on graduate coursework. These questions examined the role of core and elective courses in shaping physicist-like thinking, including how pacing, structure, content coverage and assessment practices supported or constrained students' thinking and learning process. Because U.S. physics graduate programs place heavy emphasis on completing and passing core courses, these questions and the resulting theme capture a central component of students' academic lives. This theme highlights how students experience graduate courses not only as academic requirements but as environments that can either cultivate or inhibit the kinds of reasoning valued in the discipline.

4.4 Role of graduate courses in LTP

This subtheme focuses more specifically on how core and advanced graduate courses shape students' ability to think like physicists. Students identify aspects of courses that either foster or hinder the development of deep conceptual reasoning and professional physicist habits of mind.

*4.4.1 Emphasizing conceptual understanding*

Three participants reflected on the role of graduate courses in fostering conceptual understanding and the ability to think like a physicist.

Sophie explained, "If we're talking about the E&M course, there probably were a lot of concepts that could have been learned, I would assume there are interesting things that are relevant.

I didn't get that right after taking the course. I didn't really understand what was important about it. If you know what's important about the material you're learning, then you can fit it into your framework for future problem-solving, which helps in LTP. We're just kind of collecting knowledge, and concepts are the parts that you need to actually keep with you. You can always look up how to do math. But I feel the concepts in some classes could be stressed better, stressed in a way that you may use this knowledge in these scenarios in the future."

Noah similarly highlighted the importance of connecting concepts to mathematics, noting differences across courses, "I think that it was a lot more clear, the connection between the concepts and the math in dynamical systems. I think it was just because the professor would be like, okay, you know this is why we should be doing this. And they were a lot clearer about that than in the E&M course. And again, I do think it's probably a time thing. But sometimes they [E&M instructor] would just be like, okay, here's the problem, and we're gonna start solving it and then they'd start writing on the board and not be really explaining the conceptual reason why they're doing the math they're doing. And I think that it was an assumed thing that we would know, but I certainly do not [know why they did certain math]."

He further emphasized the importance of guidance and support in developing conceptual reasoning stating, "I think focusing on the concepts would really help. And I think a really tough thing would be, being more willing to meet the students where they are, being more accepting. If someone is not so sure about a given concept, it's pretty brutal. If they're talking about a concept that you quote unquote should know, you don't want to be the one to be like, I don't know this at all. So, I think it would help better if there was more willingness towards [helping us in] understanding concepts. Concept application…would be better developed in those core courses. If we're giving an exam [with] questions that are related to concepts rather than this hard math, I think that would help develop students' LTP in terms of conceptual reasoning."

Highlighting a positive example, Brianna added, "There is one course in particular that I think did it very well, and I'm trying to compare that with the other courses that I feel are a little too focused on the mathematical side and are not really helping the thinking like a physicist mindset. Well, it's the dynamical systems [classical mechanics] course that I took here, and it just felt more like a conversation than lecturing. Things were not so abstract; I feel like breaking it down to where you're really isolating the concept and not mixing it together with too much high-level math. I think that's a way that can help you in LTP. Of course, you need math. And it's a graduate course so we are doing higher level math. But if we first start with isolating the concept, I think that helps [in LTP]."

These reflections show how graduate students navigate multiple modes of reasoning in physics courses. Their focus on connecting ideas, applying concepts and making decisions reflects Price et al.'s [15] and Robbins et al.'s [57, 58] view of authentic problem-solving. Brianna's emphasis on interactive, conversational learning echoes Ulriksen's enculturation perspective [59], highlighting how social context and disciplinary norms shape students' understanding of what it means to think like a physicist.

*4.4.2 Assessments that foster deep thinking*

All participants discussed the influence of assessments in the development of the ability to think like a physicist.

Sophie reflected, "My initial thought is that sometimes the way that work is graded stresses mathematical accuracy and not problem solving. If we're trying to genuinely learn concepts and problem-solving skills and thinking like a physicist, then I feel we should be getting better feedback or better grading on assignments where we're applying our skills. Because if we're just getting zeros or low points for small mathematical mistakes but the concepts were still good and appropriate, that's negatively reinforcing things, right? It's making [you] want to go look up how to do the math correctly to get more points rather than focusing on the genuine understanding of the situation and of the problem. And I definitely did feel that, in my core courses, in E&M and dynamical systems, there were some problems that were so difficult that people just didn't want to think about it anymore and didn't want to learn from it. They just wanted to get the points and move on. And so, I think, if the grading was structured in a way that rewarded conceptual problem-solving more than just the right numerical answer, that might encourage us to actually sit and think a little bit more and learn those skills over relying on just math."

Grace reflected that instructors are often unaware that solving a problem is not about speed. She reflected, "I feel it would be nice if instructors were aware of how it's not really about the time that you spend on solving a problem, because whenever they give out exams, and expect you to solve it in two or three hours, I've noticed that, I become limited. And so, I can give you this example [of] the quantum course that [had] really difficult and interesting questions that were like a mixture of concepts and quantitative reasoning. What I noticed was that we had all week to work on those questions. And, I realized that if I had the whole week to think about a problem without being under pressure of doing it fast, I could do so much better at times than other people. I could understand the concepts much better. I could explain it to other people, which was something that I had never been able to do before, because I was always like you're running out of time, or you have an exam. So, I think it would be really nice if the core course showed students that it's not really about how fast you are at solving problems, which I think a lot of them do when they take high-stakes exams, which is like 50% of your course grade. And they're like, okay, whatever you learned in the whole semester, you have to just blurt it out in 2 hours, and I don't think that's fair."

Ethan highlighted another aspect of improvement for core graduate physics courses stating, "You reach a boundary when you're doing research where there actually may not be a known solution to the problem that you are tackling. And I think that core courses can do a better job of showing you how to apply certain techniques to help you come closer to finding a solution to a problem that may not have a solution. Because if you're trying to find a solution to something that no one has a solution to, you think like a physicist and learn how to apply all of your knowledge to come closer to finding a solution. So, I think that could be improved on."

Brianna and Kylie contrasted courses, noting that assessments in dynamical systems promoted conceptual understanding, whereas others remained heavily math-focused or textbook-driven, Brianna said, "I think the assessments in the dynamical systems did a good job at probing the concepts that we were trying to learn. We still had to do math, but it wasn't a barrier as much as it was in E&M. I feel the assessments in E&M were far too math focused. The E&M part of it was one or two lines, and then the rest of it was just math that, in my opinion, had nothing to do with physics."

Kylie contemplated, "Maybe classical mechanics was a good course that I remember because we had good kinds of problems that we were supposed to solve, and it was helping me to develop those skills. I would say quantum mechanics was good, but it was a bit heavy in terms of the content…And then we were supposed to solve problems from the back of the textbook. So, if you can keep up with that, it is good, but if not, then you feel like you're lagging."

These reflections suggest that concerns about grading that emphasizes quantitative procedures over learning physics concepts align with Modir et al.'s [39] distinction between conceptual and algorithmic knowledge, highlighting the need for assessments to incentivize and support graduate students in LTP. Similarly, valuing problems that promote conceptual synthesis as well as explore authentic unsolved scenarios reflects Price et al.'s [15] and Robbins et al.'s [57, 58] view emphasizing authentic decision-making. Participants also noted that core courses should better expose students to the uncertainty inherent in research, where progress often involves applying tools to approach problems without known solutions. These insights suggest that assessments rewarding conceptual reasoning and reflection on authentic problem-solving process [57, 58] may help cultivate physicist-like thinking.

*4.4.3 Advanced courses enhance physicist thinking*

Five of the graduate students who had taken advanced level courses expressed their views on the role of these courses in enhancing physicist thinking.

Sophie explained, "I think the courses beyond the core courses are the ones that are much better at helping me develop the skills I need to think like a physicist, but also to think like a physicist within my subfield. There you can only learn so much broad knowledge and then you have to narrow the scope a little bit to what you like, [and] what kinds of problems you want to be solving. So, those courses taught me good conceptual things that I can use to do future research within my field as well as just gave me some base knowledge about important things…in those courses, I was taught more than memorization type things…also conceptual frameworks for how to solve problems. You have to have some conceptual knowledge and some genuine facts that you can remember about your field. But both those things I got very good practice with and knowledge of in my elective courses."

Noah highlighted how electives such as computational physics and quantum optics expanded both the conceptual and mathematical sides of their learning, "Thinking about the computational physics course, I think that was helpful especially because it was a lot wider ranging. And I did the quantum optics course, that was okay. They talked about reasonings behind doing things a lot and they talked about concepts. And they talked about the math behind it a good bit too. So, I would say those definitely were helpful [to LTP]."

Ethan also emphasized the value of electives but expressed a desire for more opportunities to engage in specialized coursework, "Yes, I do think that they are helping me [LTP]. Now, becoming a senior grad student, I think there's so much more that I still want to learn and haven't had the chance to…you can have courses that I feel could be taught endlessly. So…I would love faculty members to do more special topics…I think that that would be really helpful."

These reflections show that advanced courses play an important role in blending concepts with math and provide more opportunities for thinking and enhancing the process of LTP.

4.5 Course pacing, its impact and suggested changes

Here students reflect on the pace of graduate core courses and how it affects their learning. Many describe feeling rushed through content, limiting opportunities for deep engagement. This code also includes suggestions for pacing or structural changes that would better support LTP.

*4.5.1 Rapid pacing limits conceptual understanding*

All participants highlighted the pace of graduate core courses as a critical factor affecting the process of LTP, with excerpts illustrating this concern.

Sophie reflected on the pace of graduate core courses, explaining, "I think if any of them did move too fast, it probably was E&M. It is just a notoriously difficult course, because there's so much you can talk about. I think they moved fast…we didn't really get to sit with the concepts. And I feel, especially at the graduate level, we have a good amount of base knowledge that we probably should be stressing, more understanding big concepts. Even if that means we go slower and not learn as much. I think that would be more useful than trying to jam a bunch of things into the course that aren't necessarily that conceptually interesting or important. So, some of them did go too fast for my understanding. And then, because of that, I didn't understand much, and I didn't get much from the course. If the things are rushed, then you're just struggling to stay afloat in the course rather than genuinely trying to construct your knowledge."

Grace shared a similar experience stating, "I think, without any exception, I could say all of the physics courses that I have taken towards the end of the semester, the instructors somehow realize that they haven't covered all that they wanted to cover, so they just speed up and they try to catch up on all of the topics that they wanted. So always towards the end of the semester, I'm kind of lost and some of these instructors changed their whole teaching style. I don't find that helpful and I feel like that definitely hurts that physicist mentality that you might have, because now it's not really about understanding the concepts and being able to solve problems. But it's more about surviving the course. Let me just go over all of the topics at a very shallow level, so that I can just do well on the exam. And it just becomes completely superficial." She further gave an example of a course, "The statistical mechanics course was definitely one of those courses that started out at a slower pace and kind of accelerated towards the end, which led me [to worry] towards the end of the semester. I was more like, oh, I need to get this done. So, I went more towards taking a quantitative approach, maybe sometimes even algorithmic problem solving, without really understanding what's happening."

Noah gave examples of two other graduate courses, "I think the best was probably dynamical systems. It went at a pace that was seemingly good. I mean, maybe there's just less to cover. It went at a good pace, and then I felt like I left with understanding pretty much every class. I would leave with understanding why we needed to apply given equations. For the worst, I do think it was E&M. Personally, I've always struggled with E&M. So that makes it harder on its own. And then, like I mentioned, there's always a general hand waving like, oh, you guys know this from whenever and I think that's probably a result of needing to get through lots of content. But I think that there was just so much content, and it was moving at a bit too quick of a pace, and that caused me to not be able to think like a physicist for that course

Noah also pointed to pacing as a barrier, "I think that it definitely affected my LTP, because I had to adjust my goals from being why is this happening to being like I just need to develop a way to get the answer or find ways to be correct for assessment [to get good grades]. I think that the instructor for the graduate quantum mechanics has forgotten the experience of being a student, and maybe that led them to not recognize that they were moving a little quickly. I think the pacing is probably the biggest thing where it's tough, especially for E&M, because there is so much...And I guess, thinking about quantum mechanics, it probably had a similar situation to E&M, where there's just so much content that you have to get through and I'm hesitant to say that both should be spread across two semesters, because that would be a lot. But it's just a time thing, I mean it always is in grad school, I guess."

Ethan also echoed these concerns, "I think that some of the core courses moved way too fast and it affected my LTP, because again, I felt like I was overwhelmed with work. I didn't have the time to appreciate the skills that I was developing. I was just trying to make it week by week to submit all of the homework rather than really learning anything. I mean, I obviously did learn something, but I guess the sensation of being overwhelmed was so significant that I maybe could have learned more had the pace of the course been different."

These reflections show that insufficient time to engage with concepts can push students toward algorithmic and procedural thinking. They also highlight that meaningful reasoning requires time to think deeply and select problem-solving approaches, connect representations and evaluate solutions. Together, these insights suggest that course pacing can significantly influence students' cognitive engagement and the development of expert-like thinking in physics.

*4.5.2 Difference in levels of instructor awareness and responsiveness*

All the participants reflected on instructors' awareness of the challenges posed by demanding courses and how the instructor tackled those issues.

Sophie described how some instructors demonstrated awareness of the challenges posed by demanding courses like E&M, "I think some instructors, maybe, are aware of it, especially with E&M. We all know that that course is very long and tough. So I think some instructors understand that, and do try to be a little more lenient with various parts of the course, maybe just giving extensions on homework, or skipping certain topics in lecture because if you're starting to get overwhelmed with how much we're talking about in the class, I think it's okay to cut one topic here and there. So, I appreciate when professors are willing to do that. I think some professors care too much about tradition and being like, no, we're gonna talk about this because it's tradition, and it's like, well, if no one [in the class] cares and no one's learning the concepts about it, then we probably shouldn't talk about it, because it's not helping us learn."

Grace doubted instructors' unawareness stating, "I think they're always aware of it, and unfortunately, I don't think they do anything about it. I've seen that sometimes instructors are like, oh, I'm not gonna take more questions today because I don't wanna become rushed. Unfortunately, [this is] something that happens a lot especially in science and I guess physics. But I feel, regardless of your experience in teaching or no matter how many years you've been in the field, I would expect a good professor to have a lesson plan to know what they're gonna do in each class and teach. But a lot of times, it happens that physics instructors are like, okay, I'm just gonna go to the class and I'm gonna teach students. And, so they don't know if it's gonna work out in time. So that's usually why, towards the end of the semester, they're like, I need to go faster, I have to cover all of these concepts. So, I think, they don't usually do anything about that unless, like this quantum course that I'm talking about, the instructor had a whole book written by hand…But he knew that from this, like today, we're gonna cover pages one to 10, and then the next session, we're gonna cover pages 10 to 17. So, he had this whole thing planned out. And it really works well, because he knew from the beginning what concept he's gonna go through."

Ethan added that awareness varied and sometimes fell short despite clear signs from student performance, "I think my instructors were aware of it potentially. I think there were times when feedback was given to an instructor, and sometimes that feedback was listened to, other times it was not. And I think this is a known thing in most physics departments that some instructors don't do a very good job teaching, and that can be reflected in the overall exam scores. I've been in quite a few classes in my academic career where the overall average for an exam may be below 50%, sometimes even below 40%. And the instructors' reasoning is, students are the problem rather than I may be doing something wrong. So…it has been the case…where everyone does poorly, and at that point, the instructor needs to reevaluate, in my opinion, what they were doing in that course."

Thus, participants noted that some instructors adjusted pacing or emphasized essential topics, while others failed to respond effectively to student needs. The graduate students emphasized that what was important for developing LTP was not lots of materials thrown at them but opportunity to think deeply and learn to integrate physics and math coherently while solving complex problems. These observations align with Price et al.'s [15] and Robbins et al.'s [57, 58] focus on authentic problem-solving, showing that well-scaffolded instruction is key to fostering deep conceptual engagement.

*4.5.3 Reorganization of course content*

Four participants emphasized that not all course content is equally valuable and suggested focusing on conceptually important topics rather than trying to cover everything.

Sophie reflected, "I think some topics aren't that conceptually useful, …or just things that don't come up very often in research. So, it might be better to not do that, and to instead focus more on something that is a conceptual framework that you will use later on in your career. I just think some of them could be slowed down by just cutting some of the material that's not important conceptually because if the point of the courses is to teach us how to think like a physicist and not just to memorize an equation from a particular scenario, then, we could definitely take a little more time to go through those concepts and cut out some of the not as important stuff. If you change the structure, it'll change the pacing a little bit. If you take some content out, then [you can] slow down the pacing. But yeah, I think the structure is probably the more important thing to address, because, like I said, some courses are taught very traditionally, and I think that traditionally is not always the best way especially if your goal is again to think like a physicist and not just to know specific things."

Brianna also emphasized the importance of prioritization, particularly in compressed courses, "One answer here would be to just cut out unnecessary...E&M, for example. It used to be two courses…and now they only have one. And so, they've tried to shove more stuff into the first one, and I think that does not work. So, I think quality over quantity for sure. You can't process things if you don't have practice on it, and you can't practice everything. So, if you have too many things to do in this timeframe, focusing on doing a few things very well instead of trying to do everything [is better]. I know it's a graduate course, but we're still learning. I think, having some time in the class where the students get to practice with the instructor…that would be helpful."

Noah suggested shifting some content to specialized courses, "I'm sure there are things that could be removed. I think the problem, especially with core courses is, we're trying to get general knowledge. We're trying to ensure that every grad student has the same general knowledge…So maybe looking at things that are more specifically applied, and considering putting them in a class, that's more geared towards certain groups in the department, like particle physicists or your condensed matter physicists etc. [would help]. But I do see the reason behind the core courses, and they certainly have a necessity for everyone. So, I'm not super sure."

However, Grace felt that the course content did not need to be changed, noting, "I think I would probably go for not changing the content. I feel in most of the courses that I have taken, there's enough content. It's just typical stuff that you need to know. But I think there should be enough time spent for each concept, so evenly distributed, based on what concept you're teaching."

Thus, although there are some differences among students in whether some core course content should be kept or removed, together, these perspectives holistically highlight the balance students see between maintaining comprehensive course content and ensuring sufficient time and emphasis on conceptually important issues to develop deeper understanding and expert-like problem-solving skills, as emphasized by Price et al.'s [15] and Robbins et al.'s [57, 58].

**RQ3. How do graduate students perceive the importance of thinking like a physicist for research and teaching?**

The third theme centered on relationship between LTP and research and teaching. These questions explored the importance of LTP for conducting research and teaching others, and students also described how research and teaching experiences in turn help them LTP. Thus, this theme reflects the bidirectional, iterative nature of LTP. Students apply their developing problem-solving and reasoning skills to make sense of unsolved research problems, while simultaneously using insights from research, advisor feedback and scholarly practice to refine and strengthen their own thinking. This recursive process in which LTP supports research and research reshapes LTP, is foundational to disciplinary enculturation.

4.6 Importance of LTP for research

This subtheme highlights the centrality of physicist-like thinking for success in research. Students discuss how importance of LTP manifests in their own research fields and whether these skills are adequately emphasized in core and advanced coursework designed to prepare them.

*4.6.1 Critical for success in research*

All the participants affirmed the importance of LTP in research. Sophie emphasized the centrality of conceptual thinking to success in physics, particularly in particle theory, "I personally think that LTP is critical for success in general in physics. But I do also think in my field of particle theory, concepts are even more relevant in being successful because that's what we're doing. We're thinking about how to come up with a new theory for something and that takes a lot more conceptual knowledge than just relying on math. But some other physics research is a little more hyper specific where you don't actually actively have to think big picture and super conceptually all the time. I'm sure we all think like physicists in our work. I just think some stress conceptual thinking more than others. But I think it is critical for my field at least because if we're coming up with new models for things, then we need to obviously understand how that can fit into what we already know. And that takes a lot of concepts."

Noah also highlighted the importance of conceptual thinking but framed it within their research on student learning, "I feel like it's pretty important for success in research to think about why students choose the questions they choose or choose the answers that they choose. It is a big part of my research right now. So being able to think about the concepts that they are thinking about and how they're potentially incorrectly applying them is pretty important. And then, being careful about how I'm applying my own thoughts and concepts as I'm thinking about that is pretty important as well."

Kylie explained, "I think it [LTP] is an important skill for success [in research] …The way we deal with problems is really important [for research] and the ability to blend conceptual as well as quantitative thing…it's definitely an important thing. But I don't think it's explicitly emphasized in the courses, because the courses are more lecture based, I don't think there is a lot of emphasis on that [LTP]."

These perspectives show that graduate students view LTP as critical for doing research, regardless of their subfield. Whether they are constructing new theoretical models or solving complex problems, participants emphasized that research continually demands and in turn strengthens their physicist-like thinking. The graduate students also implied that research is a primary space to meaningfully develop their conceptual, analytical and reflective abilities.

*4.6.2 Role in teaching and mentorship*

Although we did not explicitly ask participants about their teaching goals, three participants extended this discussion to articulate its significance for teaching as well.

Grace underscored the importance of thinking like a physicist within the context of physics education, "In physics education, I think it's definitely critical, because as physics educators, you're supposed to be motivating other students to study physics and to think like physicists themselves. And I think part of being a physics educator is that you're going to become a role model to students that you're teaching. You are supposed to be a good physicist, to be able to have that critical thinking, so that when students look up to you, they're able to just follow that correct direction."

Brianna similarly emphasized its significance for research and teaching, "I think LTP is a critical skill for success in my research. For my research, I think it's very important, because not only do I need to think like a physicist, but I need to be able to teach other people to LTP. So, I definitely think getting better at it myself will help others get better at it. Yes, in some of the courses, but not all of them, [I am learning these skills]."

Ethan shared, "Part of the motivation for applying to grad school was…building these skills so that I can go back home and share these skills with my community. And so at least in the beginning of my PhD, I really wanted to go back home and be an educator. So that was part of my goals. To be able to build some skills, become an expert in a field…do some research after the PhD in that field, and eventually, at some point, come back home [and] become an educator."

These reflections underscore that in these graduate students' view, LTP is not only foundational for research but can also enable them to use their expertise to teach and guide others, extending expert-like reasoning in educational contexts.

4.7 Teaching goals, course content priorities, and pedagogical techniques for LTP

Graduate students imagined themselves as future instructors and articulated how they would design graduate courses to foster LTP. They described balancing content coverage with conceptual depth, prioritizing active learning, collaborative problem-solving, and inclusive pedagogical practices.

*4.7.1 Modification of the standard content*

All participants described how they would modify course content if they were instructors to better foster LTP.

Sophie emphasized the need to restructure course content to prioritize conceptual learning over rote problem-solving, "I definitely would modify content. Like I was saying, I don't think standard course materials are always the best way to teach a class that is more traditional. And I think if we're trying to stress thinking like a physicist, then we should stress more conceptual problem-solving, which is not usually in standard course materials. It's usually numerical problem solving. So, I definitely would restructure or modify the content to make sure that we're trying to think more like physicists and not just regurgitate math."

Noah similarly stressed the importance of aligning learning objectives with conceptual understanding rather than surface-level coverage, "I would want the learning objectives to be aligned with understanding concepts a lot better than a lot of the current courses do where making sure that I have the ability to emphasize concepts a lot more which maybe a bit abstract. But I think being able to ask why is very vital in my mind to LTP and emphasizing how important that is. And perhaps, a lot of the time, it can be hard for students in my mind, because you only have one resource which is the professor, so may be providing extra resources. If my explanation was not good enough, then they can find other explanations that might help them elsewhere in order to help them learn [to think] like a physicist."

Brianna agreed that covering all of the material should not be the top priority and suggested adjusting the course structure in response to students' needs, "I don't think I would prioritize covering all of the material. I think I would modify, and it could be modified as we go through it, depending on what the class needs. So, if the class needs more time on a certain concept, then I think we should take the time to slow it down a little bit and go a little more in depth or have more practice so that they LTP."

Ethan discussed balancing foundational knowledge with flexibility to meet students' research interests, "I think I'm more open to modifying the content to support students. Whether that's the faculty members that oversee the requirements for the PhD, I think we should have a discussion about flexibility for courses depending on the interest of the students. I feel that physicists do have more foundational knowledge compared to other disciplines. And so, it's still important to emphasize that foundational knowledge. But I imagine that there are certain aspects of a particular course that can be more flexible to meet the research interests of the students. So that would be my approach."

These proposed instructional strategies align with Modir et al.'s [39] emphasis on distinction between conceptual and algorithmic knowledge, showing that restructuring courses to emphasize key concepts and adapt to learners' needs can promote expert-like thinking.

*4.7.2 Use of various pedagogical techniques*

All participants discussed strategies and pedagogical approaches that could help students learn to think like physicists.

Sophie emphasized the value of active learning over traditional lectures, "I think, especially if the course is supposed to be about teaching us to think like a physicist, then that could be a good course to use some pedagogy of active learning, as opposed to lecture. So, it could be good to have an active learning course where you construct knowledge and groups to solve these complex open-ended problems. Just come up with techniques together and share with the class and learn from each other about different ways to approach problems. So definitely, maybe some group work and some group discussion, because I also do think that conceptual knowledge is built from talking to others as well. I think in general; knowledge is built by talking to others. But math is something that is not as opinionated. You can look up how to do a math procedure. But when it comes to approaching a problem, there can be opinions on how to best do that. So, I think incorporating some discussion about approaching problem-solving would be really good to learn from each other."

Noah highlighted the role of questioning and dialogue in guiding conceptual understanding stating, "I think there are pedagogical techniques that can help students in LTP. Maybe I don't know them all. But I'm thinking about Socratic questioning where you ask students questions in order to understand what their thinking is. And trying to guide them using that would definitely be helpful to get them to LTP. So, I think having conversations with them about their thinking would

be very vital for that and guide their thinking more along the conceptual than the math would be important."

Brianna focused on structured peer interaction and instructor feedback, "I know some version of the Think-Pair-Share, giving people a problem and letting them try to work on it on their own and then work with their peers, and then get feedback from the instructor. So, everyone goes over the problem and the instructor not only just works out the problem but maybe highlights places where the instructor knows that a lot of people will maybe fumble. So, I think that definitely is one I would use to help. People think like a physicist because it helps them not only identify the correct way, but also [identify] other routes people might have taken and maybe someone tried to invoke one concept that wasn't relevant but that also helps us to structure our physics knowledge. And so, I think that's the one I would use to help students LTP among others."

These insights show that graduate students value strategies that engage students in discussion, reflection and guided problem-solving, fostering deep conceptual understanding and thinking like a physicist.

*4.7.3 Connecting learning to research and real world*

Three participants shared their thoughts on designing graduate courses that encourage students to think like physicists and engage meaningfully with research.

Sophie reflected on the idea of a core course and its limitations, suggesting a focus on research skills, "If we're just thinking about a course that everyone should take essentially, I think it would be nice to have a course that is about research methods. There are a lot of little things that go into doing research that are not stressed at all in the core courses. The core courses are still just telling us things and it's not as much having us practice applying skills…maybe just giving students very broad questions and having them conceptually come up with ideas of how to answer them. And how to do the research to answer them. Because that's a big thing for us in reality, we have to find information from other sources to be able to do our research. So, doing some sort of investigative problem-solving for some broad generic problem that we should be able to understand with our prior knowledge from undergrad and everything, that would be interesting. Just to have goals of being able to come up with a conceptual path of how to solve a problem without necessarily having to do the math. So, a broad research application course would be interesting."

Grace emphasized the importance of connecting course content to real-world applications and helping students see beyond individual problems, "I feel like if I was teaching a graduate course, I would definitely focus on having more conversations with students about the applications of whatever we're studying. Because, at least for me, and I know that it happens a lot for other students that when you learn a concept or whenever you're solving a problem, you could think, okay, why am I doing this? And I think a good course, or a good instructor would help you see beyond whatever it is that you're doing at that moment. And I think that's also what it needs to LTP. Because you're not just focused on this one problem, I have the right answer and I'm done. You're trying to, at least in a way…come up with all these new ideas, you're solving problems that are actually related to solving actual problems. For example, when students hear about quantum computing and information, what does that mean to them? Is it just bunch of words there on paper, or bunch of bras and kets on the board or are they also getting to know about how this is actually out there in the real world. So, I guess that's kind of the vision that a physicist hopefully has. And that's kind of what I would want to have students learn as well."

These perspectives highlight that graduate students value courses that integrate conceptual reasoning, research skills and real-world connections to foster expert-like thinking.

## 5. Summary and Conclusion

Our study investigated how physics graduate students define and experience "thinking like a physicist" and how coursework and other types of professional pursuits such as research connect synergistically with the development of LTP. When investigating how students develop physicist-like thinking through the lenses of Modir et al.'s [39] epistemological framing, Price et al.'s [15] and Robbins et al.'s [57, 58] framework emphasizing authentic problem solving, and Ulriksen's framework elucidating students' navigation of disciplinary norms and cultures, the participants' experiences reveal a multi-faceted development process.

The physics graduate students in our study represented both PER and non-PER backgrounds and varied in their cultural backgrounds and undergraduate training. Nevertheless, these graduate students interpreted the interview questions and described their experiences in graduate coursework in largely similar ways, with only occasional small differences. This similarity may reflect the shared structure of their doctoral training in which all students completed the same sequence of core graduate courses, took only a few different advanced electives, served as TAs in the department and were engaged in ongoing research. The milestones for making progress for all these graduate students are the same regardless of research emphasis. Importantly, our coding revealed a high degree of consistency across participants. In most categories, students converged in their perspectives, and we did not find cases where participants expressed fundamentally opposing views. This overall convergence suggests that, despite differences in subfields or research identities, graduate students share common experiences with coursework, expectations and departmental culture, patterns that mirror Ulriksen's framework [59] that emphasizes that students have an implicit understanding of the norms and structures of the discipline they inhabit and they are shaped by them.

Across participants, a consistent theme was that LTP requires a dynamic and coherent integration of conceptual and quantitative reasoning, echoing Modir et al.'s [39] emphasis on blending multiple modes of thought and Price et al.'s [15] and Robbins et al.'s [57, 58] framework on authentic problem solving. Students stressed that neither mathematical facility with procedures nor conceptual intuition alone is sufficient. Instead, authentic physics practices demand moving fluidly between the two. Their descriptions of research pertaining to authentic unsolved problems require creativity, iteration, interpretation and refinement and align closely with Price et al.'s [15] and Robbins et al.'s [57, 58] characterization of authentic problem solving as non-linear, reflective and grounded in sense-making rather than algorithmic execution.

Instructional design emerged as a powerful influence on LTP. Participants highlighted the role of course structure and pacing in shaping physicist-like thinking. Some graduate core courses, particularly E&M, were often described as fast paced and mathematically dense, providing little room for conceptual grounding or reflective engagement. Graduate students explained that high-stakes, timed assessments limit opportunities for reflective thinking, making them feel overwhelmed and constrained by fast-paced courses. Taking inspiration from Ulriksen's framework [59], these structures shape how students understand what counts in the discipline and that can inadvertently signal that speed and facility with mathematical manipulation are more valued than deep understanding of physics.

These findings build on prior interviews with graduate students and faculty about core graduate courses [41]. The interview questions in this study were designed inspired by the earlier work, and the consistency between the two sets of findings show that current graduate course structures risk becoming missed learning opportunities rather than spaces for productive disciplinary development. Across both studies, students described the same tensions such as overly broad content coverage, limited time for sense-making, and a grading system that emphasizes correctness of complicated mathematical procedures over conceptual reasoning and LTP. In contrast, graduate students' recommendations such as pacing courses to allow for conceptual engagement, prioritizing conceptual understanding over exhaustive coverage, incorporating structured opportunities for meaningful discussion and problem-solving, and providing feedback that values understanding reflect an instructional environment more consistent with supporting disciplinary enculturation and LTP. These proposed changes echo the findings from the earlier interviews [41], collectively pointing toward a vision of graduate core courses that better support identity development and the cultivation of physicist-like thinking.

Elective courses, research experiences and teaching opportunities were described as more effective for developing and harnessing the ability to think like a physicist. These spaces offered opportunities for exploratory reasoning, collaboration and time-intensive sense-making, all central to Price et al.'s [15] and Robbins et al.'s [57, 58] authentic problem-solving framework. They also helped students negotiate their place in the discipline, resonating with Ulriksen's framework [59] that students make sense of physics and their role in it, through interactions, expectations, socialization and established forms of participation. Finally, participants framed their development as both a professional and personal process. They described their development not only as intellectual growth but also as part of becoming the kind of person who can guide others in their physics learning pursuits. In particular, several noted that LTP not only prepares them for research but also shapes their teaching and mentoring. By developing these skills themselves, they are better equipped to guide others in learning to think like physicists.

Taken together, these findings suggest that developing physicist-like thinking is a multi-faceted process influenced by coursework, pedagogical practices, research experiences, and evolving disciplinary identities. Consistent with Modir et al. [39], Price et al. [15] and Robbins et al. [57, 58] and Ulriksen's [59] synergistic frameworks, graduate students benefit most from environments that support conceptual clarity, provide opportunities for authentic problem solving and create pathways into full participation in the disciplinary community that is supportive.

Our findings point to several actionable changes physics departments could consider across the broader landscape of graduate education. Students' development of physicist-like thinking is shaped by the culture of the program, the quality of interactions with faculty and peers, and the structure of their research experiences. Creating a departmental culture that explicitly values conceptual reasoning and collaborative sense-making can reinforce these skills across multiple contexts not only in the classroom but also in research group meetings, informal discussions, and mentoring relationships. Departments might therefore consider fostering structured opportunities for intellectual exchange, such as peer-led discussion groups, interdisciplinary seminars, or collaborative problem-solving spaces where students can articulate ideas, test reasoning and learn from one another. Research advising practices that include regular conceptual check-ins, encouragement of independent questioning, and explicit modeling of expert reasoning can further support this growth.

While rebalancing course pacing, integrating sense-making activities, and adopting assessments that emphasize reasoning remain important, these changes are most effective when embedded within a graduate program that consistently prioritizes depth of understanding, intellectual curiosity and reflective practices. Any restructuring of graduate programs inevitably involves trade-offs, e.g., reducing topic coverage can raise concerns about preparation for graduate research and slowing core course pacing may extend the period of foundational training. Nevertheless, students' perspectives suggest that cultivating a culture centered on conceptual rigor, meaningful interactions, and supportive research environments is synergistic with their ability to think like physicists and prepares them more fully for both research and future teaching.

## 6. Limitations and Future Directions

This study has some limitations that should be considered when interpreting the findings. First, the data were collected from a relatively small sample of graduate students at a large public university. As such, the sample may not be representative of the broader population of physics graduate students. The participants were at different stages of their graduate programs, ranging from the second to the sixth year, and they belonged to diverse research areas. Because we only had one or two students from each year, it is difficult to determine whether perspectives on LTP differ systematically across program stages or whether there are commonalities that persist throughout graduate training. Moreover, interviews were conducted in the context of U.S. graduate physics education, which is structured in specific ways (e.g., heavy emphasis on teaching assistantship, coursework combined with research). These structural features may not generalize to graduate programs in other countries or to less research-intensive institutions within the U.S.

Another issue is that four of the seven participants are graduate students with research emphasis in physics education research which may skew the findings toward epistemological and pedagogical aspects of thinking like a physicist. Although this is a limitation as these students are likely to bring different forms of expertise, awareness and sensitivity to questions about LTP, the findings presented here do not reflect qualitative differences between PER and non-PER PhD students in the same department.

Future work could address these issues in several ways. One direction would be to expand the study by interviewing a larger and more diverse pool of graduate students from different institutions, research areas, and years in their programs, to capture a broader perspective on how LTP develops. Extending this work to undergraduate and graduate students across different types of universities and in different countries would also help reveal how cultural and institutional factors shape this process. Another step would be to include faculty members in similar interviews, to compare graduate students' views with expert perspectives and examine where students' thinking converges or diverges from disciplinary norms based upon reflections of faculty. The subthemes and codes identified here could also serve as the basis for survey instruments (e.g., Likert-scale questions) that could be distributed across larger populations of students at different levels to capture more generalizable trends.

In addition, future studies could explore instructional interventions that explicitly target common issues in LTP discussed here, such as intentionally integrating conceptual reasoning with quantitative problems currently emphasized in graduate physics coursework. Investigating the role of identity development holistically may also be fruitful, as thinking like a physicist is not only a cognitive development but also tied to sociocultural issues and students' overall evolving sense of themselves as physicists, which has implications for persistence and career trajectories. Finally, complementing interviews with classroom observations or longitudinal studies could provide richer insight into how students enact LTP in real contexts and how their understanding evolves over time.

## Data Availability Statement

The data presented in this study are available upon reasonable request from the corresponding author due to data privacy requirements of US FERPA regulations.

## Ethical Statement

This research was carried out in accordance with the principles outlined in the University of Pittsburgh Institutional Review Board (IRB) ethical policy, the Declaration of Helsinki, and local statutory requirements. The participants provided consent for use of the interview data for research and publication, and consent for quotes to be used.

## Appendix

1. What does "thinking like a physicist" mean to you?
2. How does one learn to "think like a physicist"?
3. In what ways do you think physicists approach problems differently from other disciplines?
4. Do you feel that you already think like a physicist? Explain.
5. Do you feel that your physics courses help you learn to think like a physicist? Why or why not?
6. What were your goals for joining the physics Ph.D. program? Is there any relation between your goals and becoming better at thinking like a physicist? Explain.
7. Do your goals align with how the graduate level core courses are structured and taught? Explain.
8. In what ways can graduate core courses help you learn to think like a physicist better? Explain.
9. Are there any specific skills that you think are missing or not emphasized enough in these core courses that can help you learn to think like a physicist?
10. Did you feel that any of your core courses moved too fast for your understanding? How did that affect your learning to think like a physicist? Do you think your instructors were aware of it? Did they do anything about it?
11. If you could change the pacing or structure of physics core courses, what would you adjust to help you learn to think like a physicist better?
12. Which one do you consider as your best/worst graduate core course in relation to helping you learn to think like a physicist? Why?
13. Do you feel that the courses beyond the core physics courses are helping you develop the skills needed to learn to think like a physicist? Why or why not?
14. Is learning to think like a physicist a critical skill for success in your physics research? Are these skills emphasized in your coursework at core course and advanced course levels [deliberate repetition]?
15. If you had the chance to design a graduate core course,
    a. What would be your learning objectives and goals for your students to help them learn to think like a physicist?
    b. Would you prioritize covering all the standard course materials, or would you modify the content to support students in learning to think like a physicist? Explain.

  c. Can pedagogical techniques help your students learn to think like a physicist? What pedagogical techniques would you use to help your students learn to think like a physicist? Explain.
16. Over the course of your graduate program, has your perception of being a physicist or thinking like a physicist changed? If so, how?

**References**

[1] Van Heuvelen A 1991 Learning to think like a physicist: A review of research-based instructional strategies. *Am. J. Phys.* **59** (10) 891-897.
[2] Starita J T, White G D and Sikorski T-R 2020. What makes a person a physicist? Learning Assistant and physics major views. Physics Education Research Conference Proceedings 509-514.
[3] Chasteen S V, et al. 2012 Thinking like a physicist: A multi-semester case study of junior-level electricity and magnetism. *Am. J. Phys.* **80** (10) 923-930.
[4] Singh C 2002 When physical intuition fails. *Am. J. Phys.* **70** (11) 1103-1109.
[5] Maries A, Sayer R and Singh C 2025 Investigation of student and faculty problem solving: An example from quantum mechanics. *Am. J. Phys.* **93** (6) 492-498.
[6] Tzanakis C 2016. Mathematics & Physics: An innermost relationship. Didactical implications for their teaching & learning, Montpellier, France 79-104.
[7] Uhden O, et al. 2012 Modelling mathematical reasoning in physics education. *Science & Education* **21** (4) 485-506.
[8] Karam R, Uhden O and Höttecke D 2019. The "Math as Prerequisite" illusion: historical considerations and implications for physics teaching. Mathematics in Physics Education. G. Pospiech, M. Michelini and B.-S. Eylon. Cham, Springer International Publishing37-52.
[9] Kjeldsen T H and Lützen J 2015 Interactions between mathematics and physics: The history of the concept of function—teaching with and about nature of mathematics. *Science & Education* **24** (5) 543-559.
[10] Hull M, Gupta A and Elby A 2013 How students blend conceptual and formal mathematical reasoning in solving physics problems. *Science Education* **97** 32-57.
[11] Branchetti L, Cattabriga A and Levrini O 2019 Interplay between mathematics and physics to catch the nature of a scientific breakthrough: The case of the blackbody. *Physical Review Physics Education Research* **15** (2) 020130.
[12] Palmgren E and Rasa T 2022 Modelling roles of mathematics in physics. *Science & Education* **33** 1-18.
[13] Reif F and Heller J I 1982 Knowledge structure and problem solving in physics. *Educational Psychologist* **17** 102-127.
[14] Keebaugh C, Marshman E and Singh C 2024 Challenges in sensemaking and reasoning in the context of degenerate perturbation theory in quantum mechanics. *Physical Review Physics Education Research* **20** (2) 020139.
[15] Price A M, et al. 2021 A detailed characterization of the expert problem-solving process in science and engineering: guidance for teaching and assessment. *CBE—Life Sciences Education* **20** (3) ar43.
[16] Fredlund T, Airey J and Linder C 2015 Enhancing the possibilities for learning: variation of disciplinary-relevant aspects in physics representations. *European Journal of Physics* **36** (5) 055001.
[17] Pietrocola M 2008 MATHEMATICS AS STRUCTURAL LANGUAGE OF PHYSICAL THOUGHT.
[18] Domert D, et al. 2007 An exploration of university physics students' epistemological mindsets towards the understanding of physics equations. *NorDiNa: Nordic Studies in Science Education* **3** (1) 15-28.
[19] Larkin J, et al. 1980 Expert and novice performance in solving physics problems. *Science* **208** (4450) 1335-1342.
[20] Leonard W J, Dufresne R J and Mestre J 1996 Using qualitative problem-solving strategies to highlight the role of conceptual knowledge in solving problems. *Am. J. Phys.* **64** 1495-1503.
[21] Schoenfeld A H 2016 Learning to think mathematically: Problem solving, metacognition, and sense making in mathematics. *The Journal of Education* **196** (2) 1-38.
[22] Peters P C 1982 Even honors students have conceptual difficulties with physics. *Am. J. Phys.* **50** (6) 501-508.
[23] Redish E F 2021 Using math in physics: Overview. *The Physics Teacher* **59** (5) 314-318.
[24] Bing T J and Redish E F 2007. The cognitive blending of mathematics and physics knowledge. AIP Conference Proceedings, American Institute of Physics 26-29.
[25] Sherin B L 2001 How students understand physics equations. *Cognition and Instruction* **19** (4) 479-541.

[26] Gaigher E, Rogan J M and Braun M W H 2007 Exploring the development of conceptual understanding through structured problem-solving in physics. *International Journal of Science Education* **29** (9) 1089-1110.
[27] Fauconnier G and Turner M 2003 Conceptual blending, form and meaning. *Recherches en Communication; No 19: Sémiotique cognitive — Cognitive Semiotics* **19** 57-86.
[28] Bing T J and Redish E F 2012 Epistemic complexity and the journeyman-expert transition. *Physical Review Special Topics - Physics Education Research* **8** (1) 010105.
[29] Bing T J and Redish E F 2009 Analyzing problem solving using math in physics: Epistemological framing via warrants. *Physical Review Special Topics - Physics Education Research* **5** (2) 020108.
[30] Rebello N S, et al. 2017. Transfer of learning in problem solving in the context of mathematics and physics. Learning to Solve Complex Scientific Problems. D. H. Jonassen, Routledge223-246.
[31] Eriksson M, Eriksson U and Linder C 2020 Using social semiotics and variation theory to analyse learning challenges in physics: a methodological case study. *European Journal of Physics* **41** (6) 065705.
[32] Karam R 2014 Framing the structural role of mathematics in physics lectures: A case study on electromagnetism. *Physical Review Special Topics - Physics Education Research* **10** (1) 010119.
[33] Hu D and Rebello N S 2013 Using conceptual blending to describe how students use mathematical integrals in physics. *Physical Review Special Topics - Physics Education Research* **9** (2) 020118.
[34] Meltzer D E 2002 The relationship between mathematics preparation and conceptual learning gains in physics: A possible "hidden variable" in diagnostic pretest scores. *Am. J. Phys.* **70** (12) 1259-1268.
[35] Singh C, et al. 2023. Instructional strategies that foster effective problem-solving. The International Handbook of Physics Education Research: Learning Physics, Melville, New York, AIP Publishing LLC https://doi.org/10.1063/9780735425477_017.
[36] Pospiech G 2019. Framework of Mathematization in Physics from a Teaching Perspective. Mathematics in Physics Education. G. Pospiech, M. Michelini and B.-S. Eylon. Cham, Springer International Publishing1-33.
[37] Kashyap J S and Singh C 2025 Case study examining graduate student sensemaking using the epistemic game framework for Laplace's equation in upper-level electrostatics. *Physical Review Physics Education Research* **21** (2) 020125.
[38] Justice P D, Marshman E and Singh C 2025 A Framework for Understanding the Impact of Integrating Conceptual and Quantitative Reasoning in a Quantum Optics Tutorial on Students' Conceptual Understanding. *Education Sciences* **15** (10) 1314.
[39] Modir B, Thompson J and Sayre E 2017 Students' epistemological framing in quantum mechanics problem solving. *Physical Review Physics Education Research* **13**.
[40] Montgomery B J, Price A M and Wieman C 2023. How traditional physics coursework limits problem-solving opportunities. Physics Education Research Conference 2023 230-235.
[41] Singh C and Maries A 2013 Core graduate courses: A missed learning opportunity? *AIP Conf. Proc* **1513** (1) 382-385.
[42] Campos E, et al. 2020 Students' understanding of the concept of the electric field through conversions of multiple representations. *Physical Review Physics Education Research* **16** (1) 010135.
[43] Nguyen D-H and Rebello N S 2011 Students' difficulties with integration in electricity. *Physical Review Special Topics—Physics Education Research* **7** (1) 010113.
[44] Bollen L, et al. 2017 Student difficulties regarding symbolic and graphical representations of vector fields. *Physical Review Physics Education Research* **13** (2) 020109.
[45] Hu D and Rebello N S 2013 Understanding student use of differentials in physics integration problems. *Physical Review Special Topics-Physics Education Research* **9** 020108.
[46] Maries A and Singh C 2023 Helping students become proficient problem solvers Part I: A brief review. *Education Sciences* **13 (2)** (2) 138.
[47] Singh C 2009. Problem solving and learning. AIP Conference Proceedings, AIP 183-197 https://doi.org/10.1063/1.3183522.
[48] Sirnoorkar A, Bergeron P D O and Laverty J T 2023 Sensemaking and scientific modeling: Intertwined processes analyzed in the context of physics problem solving. *Physical Review Physics Education Research* **19** (1) 010118.
[49] Singh C 2008 Assessing student expertise in introductory physics with isomorphic problems. II. Effect of some potential factors on problem solving and transfer. *Physical Review Special Topics - Physics Education Research* **4** (1) 010105.
[50] Mason A and Singh C 2010 Helping students learn effective problem solving strategies by reflecting with peers. *Am. J. Phys.* **78** (7) 748-754.

[51] Van Dusen B, Barthelemy R S and Henderson C 2014 Educational trajectories of graduate students in physics education research. *Physical Review Special Topics - Physics Education Research* **10** (2) 020106.
[52] Ghimire A and Singh C 2025 Using unguided peer collaboration to facilitate early educators' pedagogical development: An example from physics TA training. *Education Sciences* **15** (8) 1038.
[53] Colclasure B C, et al. 2024 Voices from graduate school and the workforce: identified student outcomes from completing a multi-semester undergraduate research experience capstone. *Education Sciences* **14** (6) 598.
[54] Mason A and Singh C 2010 Surveying graduate students' attitudes and approaches to problem solving. *Physical Review Special Topics-Physics Education Research* **6** (2) 020124.
[55] Maries A, Sayer R and Singh C 2020 Can students apply the concept of "which-path" information learned in the context of Mach–Zehnder interferometer to the double-slit experiment? *Am. J. Phys.* **88** (7) 542-550.
[56] Ghimire A and Singh C (2025) Perspectives of women and men students and faculty on conceptual and quantitative problem-solving in physics from introductory to graduate levels. Education Sciences **15**, 1602.
[57] Robbins M E, Davis N D and Burkholder E W 2025 Decision making in graduate physics coursework: What is being assessed versus what is expected. *Physical Review Physics Education Research* **21** (1) 010148.
[58] Robbins M E, DiQuattro G J and Burkholder E W 2025 Assessment of expert decisions in graduate quantum mechanics. *Physical Review Physics Education Research* **21** (1) 010125.
[59] Ulriksen L 2009 The implied student. *Studies in Higher Education - STUD HIGH EDUC* **34** 517-532.
[60] Saldana J (2021). The Coding Manual for Qualitative Researchers, SAGE Publications Ltd.
[61] Marton F 1986 Phenomenography—A Research Approach to Investigating Different Understandings of Reality. *Journal of Thought* **21** (3) 28-49.
[62] Marshman E, Maries A and Singh C 2024 Using multiple representations to improve student understanding of quantum states. *Physical Review Physics Education Research* **20** (2) 020152.
[63] Wawro M, Watson K and Christensen W 2020 Students' metarepresentational competence with matrix notation and Dirac notation in quantum mechanics. *Physical Review Physics Education Research* **16** (2) 020112.
[64] Maries A, Lin S-Y and Singh C 2017 Challenges in designing appropriate scaffolding to improve students' representational consistency: The case of a Gauss's law problem. *Physical Review Physics Education Research* **13** (2) 020103.
[65] Meltzer D E 2005 Relation between students' problem-solving performance and representational format. *Am. J. Phys.* **73** (5) 463--478.